\def\be{\begin{equation}}
\def\ee{\end{equation}}
\def\ber{\begin{eqnarray}}
\def\eer{\end{eqnarray}}
\def\bers{\begin{eqnarray*}}
\def\eers{\end{eqnarray*}}

\documentclass[aps,prb,twocolumn,showpacs,groupedaddressamsmath,amssymb]{revtex4-1}
\usepackage{dcolumn}   
\usepackage{amsmath}
\usepackage{multirow}
\usepackage{graphicx}    
\usepackage{subfigure}
\usepackage{comment}
\usepackage{color}
\usepackage{makecell}
\usepackage{longtable}
\usepackage{ifthen}
\newboolean{includefigs}
\setboolean{includefigs}{true}      
\newboolean{includetext}
\setboolean{includetext}{true}     
\newcommand{\condcomment}[2]{\ifthenelse{#1}{#2}{}}

\begin{document}

\title{Accurate high-throughput screening of I-II-V 8-electron Half-Heusler compounds for renewable-energy applications}
\author{ Bhawna Sahni, Vikram, Jiban Kangsabanik and Aftab Alam}
\email{aftab@phy.iitb.ac.in}
\affiliation{Department of Physics, Indian Institute of Technology, Bombay, Powai, Mumbai 400 076, India}

\begin{abstract}
 Renewable energy resources have emerged as the best alternatives to fossil fuel energy which are rapidly declining with time. Choice of correct materials plays a key role in designing devices with enhanched efficiency. Half-Heusler alloys, with various valence balanced compositions, have been phenomenal in showcasing a variety of rich properties, and thus provide a general plateform to search for candidates suitable for renewable energy applications. Here, eight valence-electron count Half-Heusler(HH) alloys have been studied using reliable first principles calculations in the search of potential candidates for thermoelectric (TE), solar harvesting, topological insulator (TI) and transparent conductor (TC) applications. The initial screening parameters used for our study are chemical and thermal stability, band gap, nature of bandgap and band inversion strength. The screening is further followed by application specific descriptor calculations to predict possible compounds for energy applications. We have performed quasistatic G0W0 calculation starting from HSE groundstate wavefunction to predict the most accurate estimation of bandgap for these class of compounds. A total of 960 compounds were simulated. 121 out of 960 compounds were found to be thermally and chemically stable. 31 compounds with bandgap less than 1.5 eV were studied for thermoelectric application out of which 13 compounds were found to show thermoelectric figure of merit $\text{ZT} > 0.7$ for both p-type and n-type conduction. 30 compounds with band gap 1-1.8 eV were studied for optoelectronic application out of which 13 compounds were found to show Spectroscopic Limited Maximum Efficiency (SLME) more than 20$\%$, comparable to existing state of the art materials. 21 compounds were found to show band inversion at ambient conditions which is a necessary condition for topological insulators. The surface band structure calculations for one of the promising candidate was done to check robustness of the topological behaviour. 29 compounds were found to have bandgap more than 2 eV which are promoted for transparent conductor applications with further band engineering. We strongly believe that our calculations will give useful insights to experimentalists for synthesizing and investigating proposed compounds for different energy applications.

\end{abstract}

\date{\today}

\maketitle
\section{Introduction}
The demand for alternative energy resources has been ever increasing due to rapid decline of fossil fuel energy resources. Renewable energy resources are the best alternatives for solving this problem and have indeed emerged as a unique solution to circumvent this. Among others, the technologies based on photovoltaics, thermoelectrics, topological electronics etc. remain at the forefront in providing state of the art solutions to renewable energies. One of the key requirement, for devices based on all these technologies, is the correct choice of materials. Hunt for novel materials, which can enhance the efficiency of these devices, has always been an active area of research.

Historically, Half Heusler compounds show a plethora of interesting properties, thus pave a general platform to search for  candidates suitable for renewable energy applications.\cite{HH_all}$^{-}$\cite{18elec1}  They have a general formula unit XYZ, where X and Y are in, general, transition metals and Z is a p-block element. They are easy to synthesize with easily tunable band gaps and can have numerous possible combinations due to many choices for the three sites. Some examples of half heusler compounds having expceptionally good properties  are  p-type TaFeSb-based HH compounds\cite{TaFeSb} which show a record high thermoelectric performance (ZT) of $\sim$ 1.52 at 973 K and hence are highly promising for thermoelectric power generation. On the other hand, TaIrGe\cite{TaIrGe} is found to be p-type transparent conductor having a strong optical absorption peak at 3.36 eV with a phenomenal high hole mobility of 2,730 ${cm}^{2}{V}^{-1}{s}^{-1}$ at room temperature. LnPtBi (Ln = La, Y) HH compound\cite{LnPtBi} was the first member found to show non-trivial topological electronic structure making these class of materials also promising  for the search of new topological quantum phases. HH compounds can be classified on the basis of sum of the valence electrons of X, Y and Z elements. The 18 valence-electron systems of HH alloys have been widely studied.\cite{18elec1}$^{,}$\cite{18elec2}$^{-}$\cite{18elec5} But, 8 valence-electron systems are less explored and hence are viable candidates for high throughput computation predictions. High throughput computation has emerged as a great tool for the discovery of new compounds for various applications.\cite{highthroughput1}$^{-}$\cite{highthroughput3} Such studies can save tremendous amount of time/resources that can go in the corresponding experimental studies, yet provide useful insights to experimentalists. 

An interesting class of 8 valence-electron HH systems are I-II-V class of compounds. There has been very few theoretical and experimental studies on these compounds. Although a few studies based on the optoelectronic properties have been reported\cite{13410}$^{-}$\cite{13418} but, a reliable/accurate prediction of bandgap still remains a concern. Yadav {\it et al.}\cite{Li_based} have studied the thermoelectric properties of Li-based half-Heusler alloys using DFT in which they studied the thermoelectric properties of LiYZ ( Y = Be, Mg, Zn, Cd and Z = N, P, As, Sb, Bi).  Jaiganesh {\it et al.}\cite{NaZnX} have studied the electronic and structural properties of NaZnX ( X = P, As, Sb) compounds.  Some of these class of compounds have been experimentally synthesized. \cite{expsyn2}$^{,}$\cite{expsyn3} Bacewiz {\it et al.}\cite{1342} have prepared and characterized LiZnP, LiCdP and LiZnAs and found them to be p-type semiconductors with the help of optical and electrical measurements. The thermoelectric performance of LiZnSb was measured experimentally\cite{LiZnSbexpt} motivated by the theoretical calculations\cite{LiZnSbth1} on a bunch of antimony containing compounds. This experimental synthesis\cite{LiZnSbexpt} yields a p-type LiZnSb in hexagonal phase. Later, Miles A. White \cite{LiZnSbth2} synthesized cubic LiZnSb and reported both n-type and p-type forms of this material to exhibit high ZT values of 1.64 and 1.43 respectively at 600 K with light doping. However, they assumed a constant value of lattice thermal conductivity which gives poor predictability of the figure of merit (ZT). Although there are few thermoelectric studies on the I-II-V class of Heusler compounds, a thorough search for promising candidates with a more reliable high throughput in-depth study is lacking. Furthermore, there are still plenty of room for identifying reliable photo-voltaic materials from this class of compounds. Topological quantum materials are other potential set of materials, many of which are expected to find its home in this class, but are hardly explored.\cite{TopologicalIntro}$^{,}$\cite{topo0}

Traditional high throughput computation based on {\it ab-initio} calculations mostly relies on the use of simple local density or generalized gradient (LDA or GGA) type exchange correlation functionals.\cite{LDA/GGAhighth} These functionals, though computationally faster, have severe drawbacks specially for band gapped systems. The shortcomings arise due to the underestimation of the band gap, as illustrated in various earlier studies.\cite{LDAprob} Linear response GW calculations\cite{Linear_responseGW}$^{,}$\cite{Linear_responseGW1} within the hybrid (HSE06) functional\cite{HSE06} comes to a rescue here, with a much better band gap predictions. HSE-GW is computationally much more expensive than LDA/GGA, but are found to provide much closer comparison with experimental findings.\cite{GWandexptcomp}
Appropriate design of descriptors for a given application also plays an extremely crucial role in any high throughput computational study.  This aspect is also loosely taken up in many previous studies.\cite{highthroughput1}$^{,}$\cite{LiZnSbth1}$^{,}$\cite{LiZnSbth2} In the present manuscript, we have carefully taken up these two aspects and efficiently design a number of reliable descriptors to screen three types of functional materials i.e. thermoelectric, photovoltaic and topological quantum materials. These descriptors are calculated based on the extremely accurate band gap evaluated from the HSE-GW simulation.

\begin{figure}[htbp]
	\centering
	\includegraphics[scale=0.32]{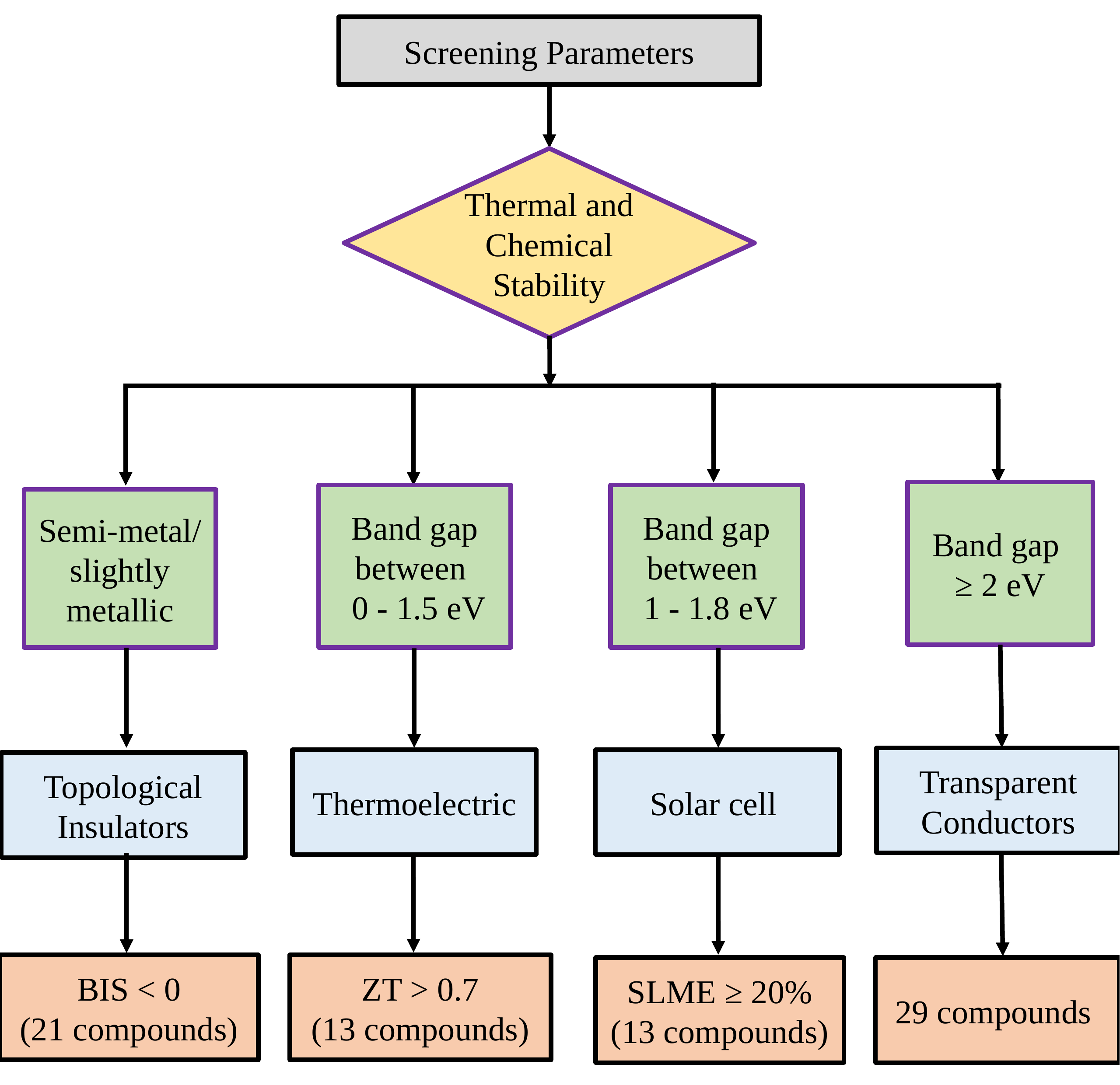}
	\caption{Flowchart explaining step-by-step screening procedures in shortlisting the compounds for different applications. HSE-GW band gap values were used to shortlist these compounds.}
	\label{screening}
\end{figure}

Here, we have performed a systematic and detailed high throughput calculations, based on the first principles simulation, to screen  I-II-V class of 8-electron half Heusler compounds which can be promising for thermoelectric, photovoltaic and topological insulator applications. A detailed electronic structure, phonon, thermoelectric, opto-electronic and topological insulating  properties are simulated. Figure \ref{screening} shows the screening criteria which we follow for the high throughput calcualtions. Initially the thermal and chemical stability of these compounds was examined. Later, the thermally and chemically stable compounds were shortlisted for different energy applications based on their HSE-GW band gap values. The parameters band inversion strength (BIS), ZT and SLME used for topological, thermoelectric and solar cell applications will be explained in the forthcoming sections.      
Such theoretical study will definitely help in discovering new  potential materials and hence motivate experimentalists to synthesize them and verify the theoretical findings.

\section{Computational Methods}
For the DFT\cite{DFT} calculations, we have used Vienna Ab-initio Simulation Package (VASP)\cite{VASP1}$^{-}$\cite{VASP3} with a projected augmented wave basis\cite{PAW basis} and the generalized gradient approximation (GGA)\cite{GGA} exchange-correlation functional of Perdew$-$Burke$-$Ernzerhof (PBE).\cite{PBE} Single-shot GW (G0W0) calculations are performed with hybrid functional HSE06\cite{HSE06} to get more accurate estimate of the bandgap. A kinetic energy plane wave cut-off of 500 eV was used. The Brillouin zone sampling was done by using a $\Gamma$-centered $k$-mesh.\cite{brullioun_zone} For all the compounds, a $k$-mesh of $10\times10\times10$ (for ionic relaxations) and $20\times20\times20$ (for self-consistent-field calculations) were used for PBE calculations. For hybrid (HSE06)\cite{HSE06} and G0W0 calculations, a $10\times10\times10$ $k$-mesh was used. In order to check band inversion for topological insulating compounds, spin-orbit coupling (SOC) was included. Cell volume, shape, and atomic positions for all the structural configurations were fully relaxed using conjugate gradient algorithm\cite{CGA} until the energy (forces) converges upto $10^{-6}$ eV (0.001 eV/$\AA$). The tetrahedron method with Blochl correction\cite{blochl_correc} was used for accurate electronic density of states calculations.

HH compounds have an interpenetrating face-centered cubic structure with a general composition of XYZ. X and Y are in, general, transition metals and Z is a p-block element.
All the compounds (XYZ) were considered to crystallize in the F$\bar{4}$3m ($\#$216) space group with X-atom at 4a (0,0,0), Y at 4c (0.25, 0.25, 0.25), and Z at 4b (0.5, 0.5, 0.5) Wyckoff positions. The formation energy ($\triangle E_F$) was calculated as:

\begin{equation}
\triangle E_F = E(XYZ)-[ E(X)+ E(Y)+ E(Z)]
\end{equation}
where $E(XYZ)$ is the total energy of the corresponding compound and $E(X)$, $E(Y)$, $E(Z)$ are the energies of the constituent elements in their equilibrium phases. A negative (positive) value of $\triangle E_F$ signifies stability (instability) of formation of the given compound with respect to its constituent elements.

The semi-classical Boltzmann transport formalism as implemented in BoltzTraP code\cite{BoltzTraP} was used to calculate the TE properties (Seebeck coefficient (S), electrical conductivity ($\sigma$) and electronic contribution to the thermal conductivity ($\kappa_e$)) of selected candidates. Density functional perturbation theory (DFPT) combined with phonopy\cite{phonopy} was used to obtain relevant phonon properties. Further details are provided in the supplementary material (SM).\cite{supl}  Lattice thermal conductivity is calculated using the Debye-Callaway (D-C) model.\cite{dcmodel} 
Within this model, the major contribution to the lattice thermal conductivity comes from the 3-phonon scattering process  (normal $\&$ Umklapp phonon scattering) and is majorly due to the 3 acoustic modes of vibrations. The thermal contribution due to each of the acoustic vibrational mode in D-C model\cite{dcmodel,Ourpaper} is given as,     
\begin{equation}
\kappa_{i}=C_{i}T^{3}/3\left\{ \intop_{0}^{\theta_{i}/T}\frac{\tau_{c}^{i}(x)x^{4}e^{x}}{\left(e^{x}-1\right)^{2}}dx+\frac{\left[\intop_{0}^{\theta_{i}/T}\frac{\tau_{c}^{i}(x)x^{4}e^{x}}{\tau_{N}^{i}\left(e^{x}-1\right)^{2}}dx\right]^{2}}{\intop_{0}^{\theta_{i}/T}\frac{\tau_{c}^{i}(x)x^{4}e^{x}}{\tau_{N}^{i}\tau_{U}^{i}\left(e^{x}-1\right)^{2}}dx}\right\} 
\end{equation}

\noindent where $x=(\hbar\omega/k_B T)$ and the index $i$ denotes LA, TA and TA$^\prime$ modes of vibration. The constant $C_i$ is given by, 
\begin{equation}
C_{i}=\frac{k_{B}^{4}}{2\pi^{2}\hbar^{3}\nu_{i}}
\end{equation}

The scattering rates corresponding to the normal and Umklapp scattering processes are,
\begin{equation}
\frac{1}{\tau_{N}^{LA}(x)}=\frac{k_{B}^{3}\gamma_{LA}^{2}V}{M\hbar^{2}\nu_{LA}^{5}}\left(\frac{k_{B}}{\hbar}\right)^{2}x^{2}T^{5}
\end{equation}

\begin{equation}
\frac{1}{\tau_{N}^{TA/TA'}(x)}=\frac{k_{B}^{4}\gamma_{TA/TA'}^{2}V}{M\hbar^{3}\nu_{TA/TA'}^{5}}\left(\frac{k_{B}}{\hbar}\right)xT^{5}
\end{equation}

\begin{equation}
\frac{1}{\tau_{U}^{i}(x)}=\frac{\hbar\gamma^{2}}{M\nu_{i}^{2}\theta_{i}}\left(\frac{k_{B}}{\hbar}\right)^{2}x^{2}T^{3}e^{-\theta_{i}/3T}
\end{equation}

\noindent where, $k_B$ is the Boltzmann's constant, $\hbar$ is the reduced Plank's constant, T is the temperature, $\theta$ is the Debye temperature, V and M are the volume and mass per atom respectively, $\gamma$ is the mode gruneisen parameter, $\nu$ is velocity and $\omega$ is the angular frequency.
In addition to lattice and electronic contribution to the thermal conductivity, we have also calculated the bipolar components of thermal conductivity ($\kappa_b$) which often becomes important at relatively higher temperature. Further details about the formulation of $\kappa_b$ is provided in SM.\cite{supl}

The optical properties were simulated from the linear response calculations as implemented in VASP. The optical transition probabilities were calculated from the dipole-dipole transition matrix elements. The independent particle approximation was used to obtain frequency dependent dielectric function. All the other optical properties are then obtained from real and imaginary part of the dielectric function. The band structure was obtained by using PBE exchange correlation functional. Because the optical properties are extremely sensitive to the band gap of the concerned material, a more accurate HSE-GW calculations were performed to estimate the same. SLME is the maximum solar power conversion efficiency which incorporates the nature of the bandgap and absorption coefficient for a particular compound.\cite{solar1}$^{-}$\cite{solar3} Details of SLME formulation can be found in our earlier article.\cite{Jibanpaper} SLME\cite{SLME,Jibanpaper} was used as a figure of merit for the photovoltaic applications.

\begin{figure}[t]
	\centering
	\includegraphics[scale=0.34,trim = 0 2cm 0 0]{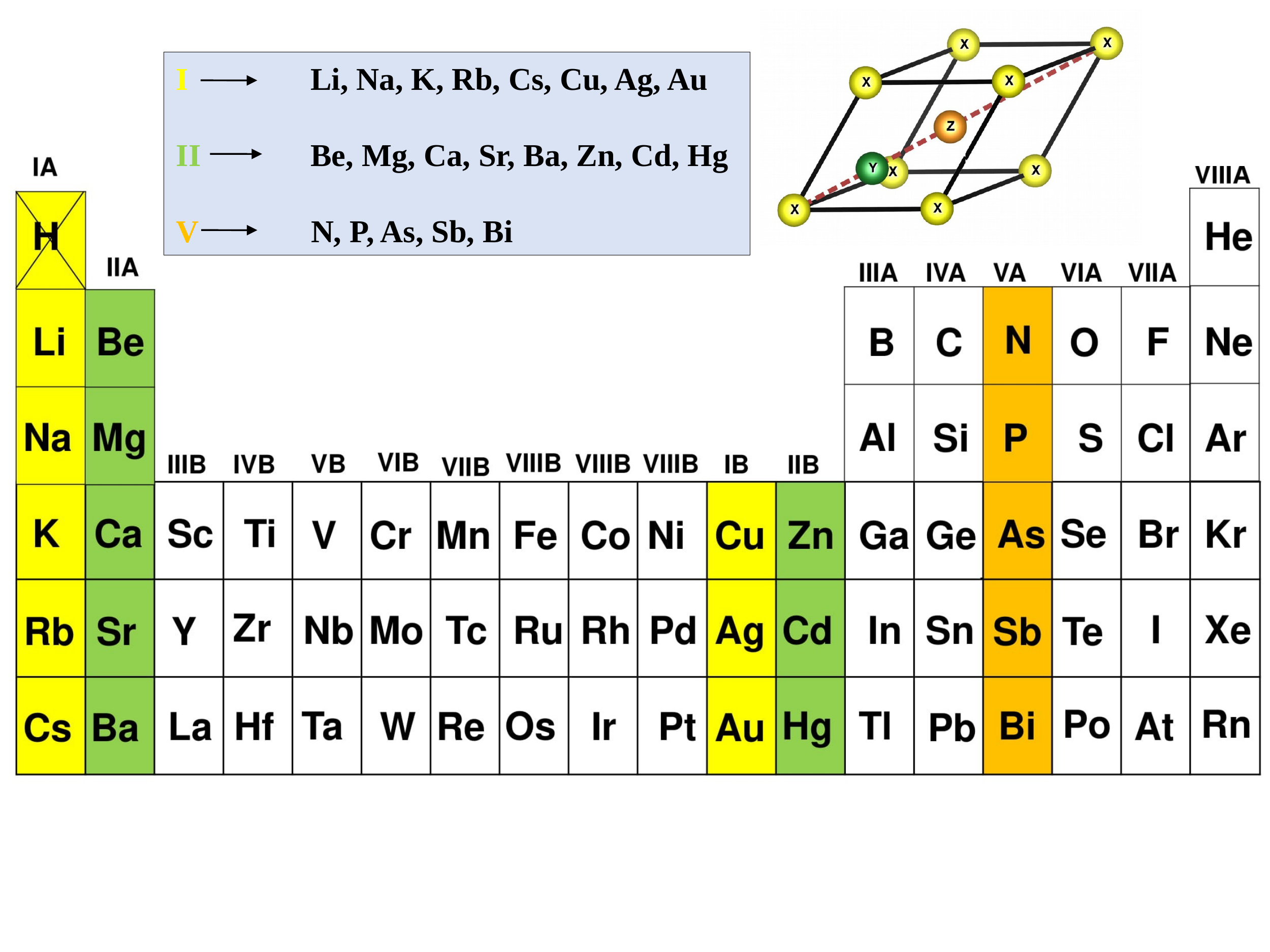}
	\caption{Periodic table of elements showing I-II-V class of elements (highlighted) used in the present high throughput study. The figure (on the top) shows the primitive unit cell of Half-Heusler compounds with one possible configuration of X, Y and Z elements.}
	\label{periodictable}
\end{figure}

\begin{figure*}[t]
	\centering
	\includegraphics[scale=0.7]{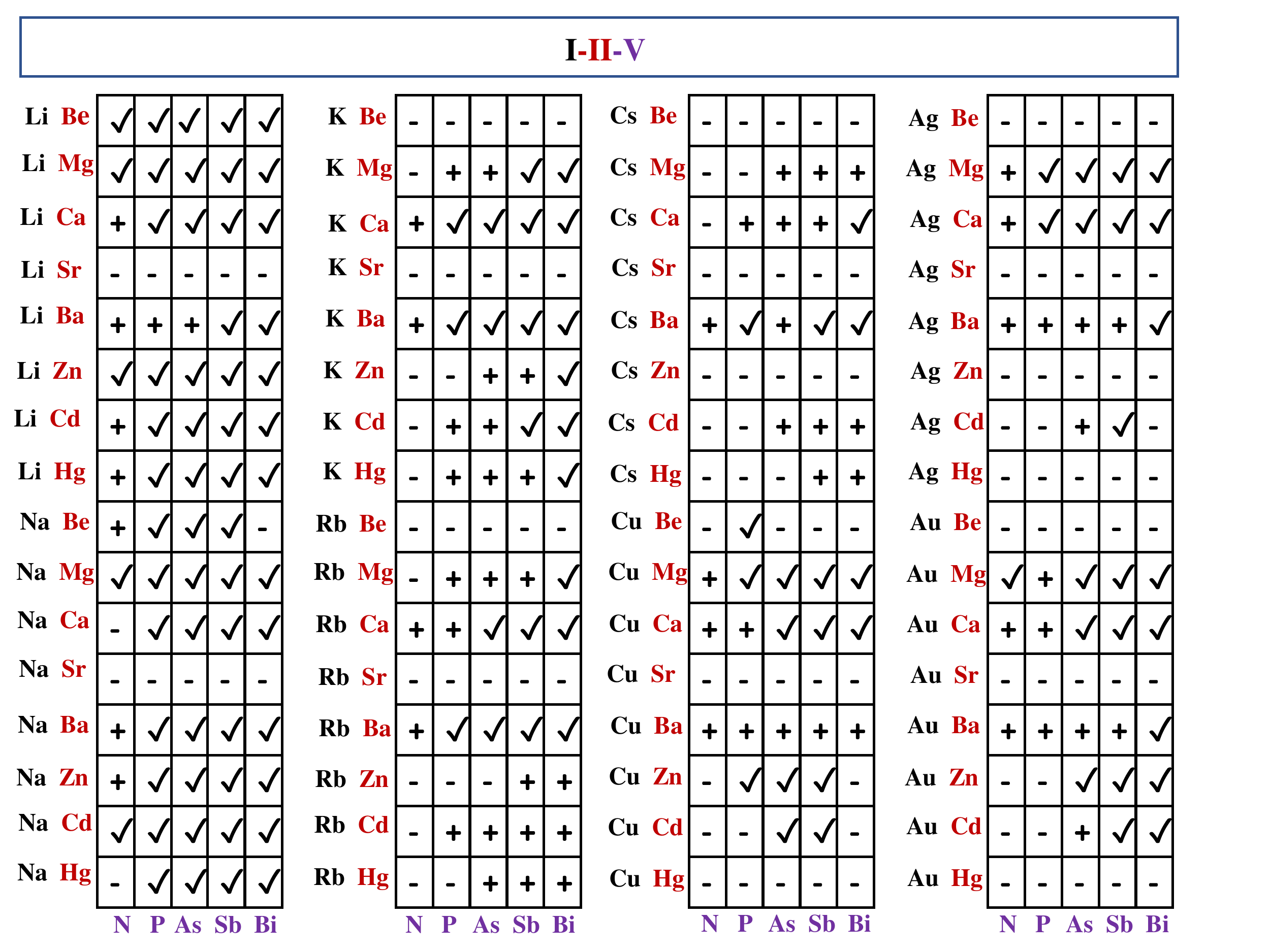}
	\caption{Chemical and thermal stability of all possible I-II-V class of 8 electron half-Heusler compounds $XYZ$ obtained by taking the most stable configuration out of the three possibilities ($XYZ$, $YXZ$, and $XZY$ for every compounds).  Negative (--) sign indicate those compounds which are chemically unstable, positive (+) sign shows compounds which are chemically stable but thermally unstable (imaginary phonon frequencies), while tick ($\checkmark$) sign indices compounds which are chemically as well as thermally stable.}
	\label{Eform}
\end{figure*}
\section{Results and Discussion}
 Figure \ref{periodictable} highlights the elements from I, II and V group of periodic table chosen for our high throughput study. The elements from VB group were not taken since for XYZ alloys to form, Z should be a p-block element.

 This gives us a total of 320 compounds to be studied. Furthermore, there are 3 unique possibilities for site preference of X, Y and Z elements at the three Wyckoff positions (4a, 4b and 4c) in a given half Heusler structure (i.e. XYZ, XZY and YZX). This makes a total of 960 (320$\times$3) simulations to be carried out. As illustrated before, depending on the concerned applications, the screening parameters used in the present study are formation energy (chemical stability), phonon dispersion (thermal stability), accurate bandgap values (PBE + HSE06 + GW), nature of bandgap (direct/indirect), thermoelectric figure of merit(ZT), band inversion strength (BIS) and SLME.

\subsection{Initial Screening Parameters}
\subsubsection{Stability}   
In this section, we have examined the chemical and thermal stability of all the compounds. For every compound, we simulated the total energy for the three distinct structural configurations (XYZ, XZY and YZX), and chose the energetically most stable one. As such, we shortlisted 320 unique compounds out of 960.

Figure \ref{Eform} shows the chemical and thermal stability of these 320 unique compounds. Here, negative (--) sign indicates those compounds which are chemically unstable (+ve formation energy), positive (+) sign indicates those which are chemically stable (-ve formation energy) but thermally unstable (imaginary phonon frequencies), while tick ($\checkmark$) sign indicates those which are both chemically as well as thermally stable.  192/320 compounds were found to have negative formation energy. Phonon dispersion for all of these 192 compounds are then simulated. 71/192 compounds were found to show imaginary phonon frequencies. Hence, a total of 121/960 compounds were found to be chemically as well as thermally stable, which are then chosen for further screening. The numerical values of formation energies and the possibility of the thermal stability for all these 192 compounds are shown in Table SI of SM.\cite{supl}

\subsubsection{Band gap}
Band gap was used as the next screening parameter to categorize the shortlisted 121 compounds into specific applications. Band gaps are initially evaluated using PBE exchange correlation functional. PBE functional, however, is traditionally known to underestimate the band gap values.  As such, compounds having bandgap between 0-1.5 eV or with negligibly small density of states at Fermi level (within PBE) were further chosen for HSE-GW calculations to get an accurate estimate of the band-gap. 21 compounds were found to be highly metallic within PBE simulations, and are not expected to open a gap under HSE-GW calculations, hence neglected.

\begin{figure}[t]
	\centering
	\includegraphics[scale=0.68]{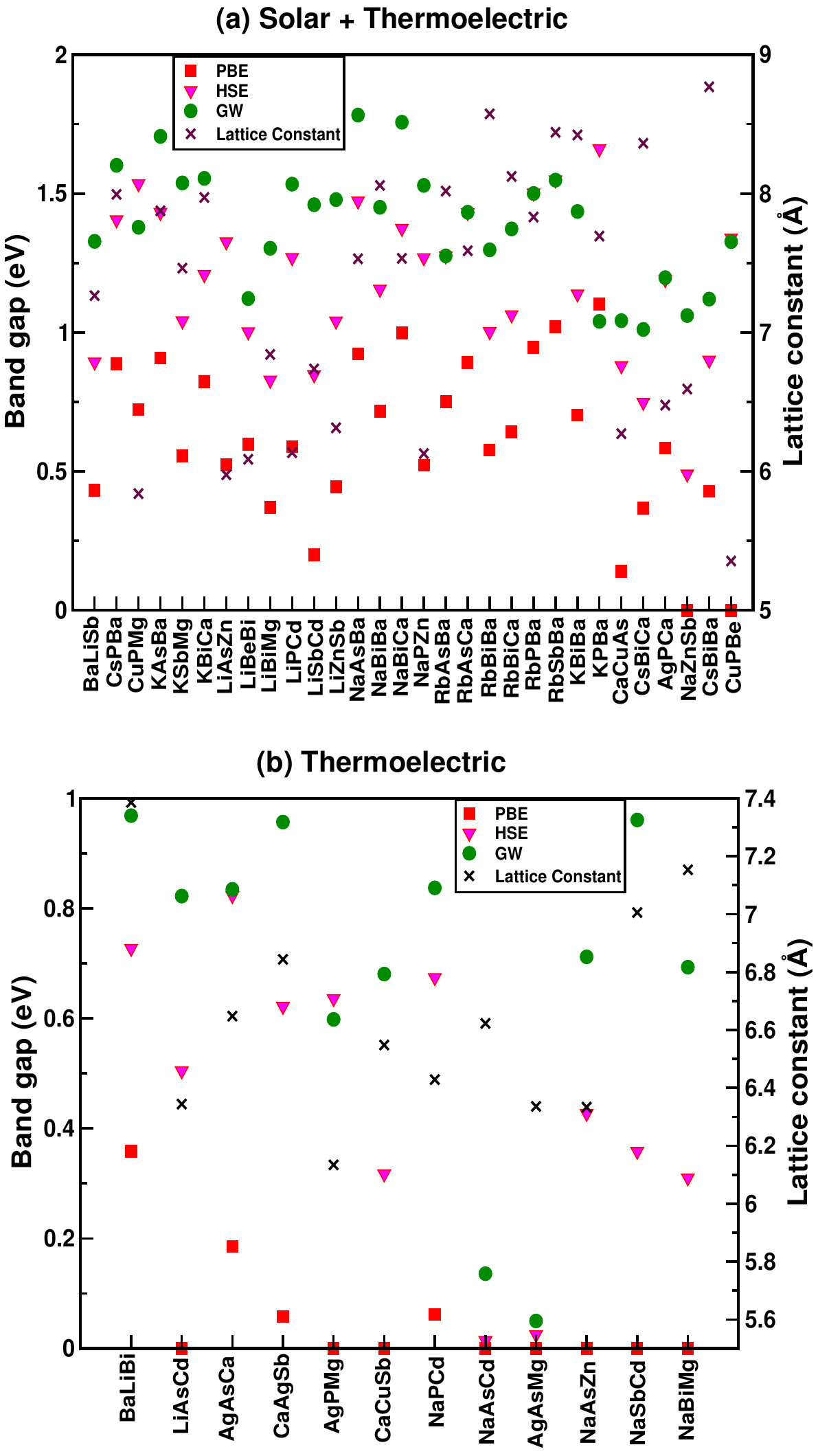}
	\caption{ Band gap data calculated at the level of  PBE (square; red), HSE06 (triangle down; magenta) and HSE06 + GW (circle; green) functionals. The absence of data for any given compound corresponding to a particular exchange correlation functional implies its metallic behaviour. Optimized lattice constants calculated at the PBE level are shown by star ($\times$) symbol. Top (bottom) panel shows compounds with HSE06 + GW band gaps in the range 1.0-1.8 eV ($<$ 1 eV).  30 compounds are found with band gap in the range 1.0-1.8 eV, suitable for solar application, out of which 19 compounds with band gap 1-1.5 eV are suitable for both solar and thermoelectric application. 12 compounds with band gap less than 1 eV are also chosen for purely thermoelectric studies.}
	\label{bandgap}
\end{figure}

A total of 31 compounds with HSE-GW bandgap less than 1.5 eV were chosen for thermoelectric studies. 30 compounds were found to show band gap between 1.0-1.8 eV which are suitable for optical absorption and hence were shortlisted for performing further simulation for photovoltaic application. Figure \ref{bandgap}(a) shows the simulated band gaps and the optimized lattice parameters for 30 compounds with band gap between 1.0-1.8 eV, found suitable for solar cell application. This also includes 19 common compounds with band gap 1-1.5 eV that are suitable for both solar and thermoelectric application.  Figure \ref{bandgap}(b) shows the simulated band gaps and the optimized lattice parameters for the compounds suitable for only thermoelectric application (HSE-GW band gap $<$ 1.0 eV). One can easily notice that PBE functional significantly underestimates the band gap. In addition, 22 compounds were found to be semi-metallic/slightly/weakly metallic, which may serve as potential candidates for topological insulator and hence further investigated.

29 compounds were found to have bandgap above 2 eV (within PBE) which is a suitable band gap range for discovering transparent conductors. Although, we have not performed detailed calculations for compounds having bandgap greater than 2 eV, these candidates can be promoted for transparent conductor applications with further band gap engineering. The list of these compounds are given in Table SIV of SM.\cite{supl} Some of these class of compounds have already been synthesized experimentally in cubic phase. Table SII and SIII of SM\cite{supl}  show a comparison of the experimental lattice parameter ($a_{exp}$) and band gap with our theoretically optimized values for those compounds which are synthesized experimentally. Our simulated values matches fairly well with the experimentally reported ones. This sets a benchmark and signifies the accuracy of our calculations.

After categorizing the pool of compounds for three different applications, we then performed detailed calculations for each of them. We will now show the simulated results for each applications and hence screen the most promising candidates out of them.

\subsection{Thermoelectric application}
Thermoelectric properties of 31 compounds with band gap $<$ 1.5 eV were simulated. D-C model was used to calculate the lattice thermal conductivity, which is a reasonable approximation for $\kappa_L$ in high temperature range. The accuracy of D-C model and cross-validation with experimental results can be found in our previous report.\cite{Ourpaper} 13 compounds were found to have figure of merit, $\text{ZT} > 0.7$ for both n-type and p-type conduction. Table \ref{ZTmax} represent the Seebeck coefficient (S), power factor ($S^2$$\sigma$), ZT, and optimal carrier concentration (N) for the compounds having $\text{ZT} > 0.7$ for both n- and p-type conduction.  The optimal concentration (N) is the carrier concentration at which ZT is maximum in the given temperature range. Similar data for all the 31 compounds are shown in Table SV of SM.\cite{supl}

\begin{table}[t]
	\centering{}
	\begin{tabular}{|c|c|c|c|c|c|c|c|}
		\hline 
		S.No. & Compound  & Doping & S & $S^2\sigma$ & ZT$_{max}$ & N & T\tabularnewline
		\hline 
		\hline 
1       & BaLiBi  & n     &  373.3   &  2.49  & 0.70   & 2.4 & 900  \tabularnewline         
         &         & p     &  386.0    &  0.96  & 1.01   &  0.9  & 900  \tabularnewline
	\hline 
2      & CaAgSb  & n     &  323.3   &  3.85  & 0.99   &  5.2  & 900   \tabularnewline       
	   &        & p     &  330.5    &  1.49  & 0.56   &  1.3  & 900   \tabularnewline 
	\hline
3      & AgPMg   & n     &  135.7   &  4.79  & 0.76   &  26.5  & 900   \tabularnewline         
	&        & p     &  337.1    & 3.88   & 1.56   &  5.4  & 900  \tabularnewline  
	\hline 
4      & AgAsMg  & n     &  174.2   & 5.43  & 0.70   & 13.3 & 900  \tabularnewline      
	&        & p     &  197.2    &  2.74  & 0.93   &  7.6  & 900   \tabularnewline 
	\hline
5      & NaSbCd  & n     &  147.3   &  3.81  & 0.78   &  12.2  & 900   \tabularnewline       
	&        & p     &  306.1    &  4.39  & 1.38   &  8.2  & 900   \tabularnewline
	\hline  

6      & BaLiSb  & n     &  384.6   &  2.09  & 1.01   & 1.9 & 900   \tabularnewline     
	&        & p     &  425.6    &  0.83  & 0.74   &  0.7  & 900  \tabularnewline 
	\hline
7      & LiSbCd  & n     &  177.2   &  5.11  & 0.91   &  12.8  & 900   \tabularnewline         
	&        & p     &  308.9    &  3.77  & 1.15   &  4.6  & 900  \tabularnewline 
	\hline
8       &KPBa  &n     &  202.27   &  6.13  &0.74   & 44.94 &900  \tabularnewline         
&        &p     &  364.46    &  3.34  &1.03   &  19.14  &660  \tabularnewline 
	\hline 
9      & RbPBa  & n     &  183.1   &  5.67  & 0.73   & 47.9 & 900   \tabularnewline     
	&        &p     &  390.1    &  2.45  & 1.37   &  10.8  & 520  \tabularnewline 
	\hline
10      & CuPMg  & n     &  246.4   & 8.01  & 0.88   &  18.3  & 900   \tabularnewline         
	&        & p     &  392.2    &  1.75  & 1.04   &  1.9  & 620  \tabularnewline   
	\hline
11      & CuPBe  & n     &  409.5   &  4.39  & 0.93   &  5.0  & 900   \tabularnewline       
	&        & p     &  442.9    &  2.64  & 0.83   &  1.8  & 900    \tabularnewline 
	\hline
12      & NaZnSb  & n     &  207.1   &  4.41  & 0.77   & 5.2 & 900   \tabularnewline     
	&        & p     &  400.1    &  2.09  & 1.04   &  1.7  & 880  \tabularnewline 
	  \hline
13      &CaCuSb  &n     &  317.3  & 3.77 & 0.92 & 5.2 & 900 \tabularnewline
&        &p      & 287.9  & 1.85  & 0.59   & 1.9 & 900 \tabularnewline \hline 	 	
	              	   
		\hline 		
	\end{tabular}
\caption{Selected candidate materials whose maximum ZT value is $>$0.7 for both n- and p-type conduction. Listed are the maximum ZT values, and the corresponding Seebeck coefficient (S, $\mu$VK$^{-1}$ ), Power factor ($S^2\sigma$, mWm$^{-1}$K$^{-2}$), carrier concentration (N, $\times10^{19}$cm$^{-3}$) and the temperature (T, K).} \label{ZTmax}
\end{table}



In the next section, we shall choose one of the promising candidate material AgPMg and discuss its TE properties in some details.


\begin{figure}[h]
	\centering
	\includegraphics[scale=1.0]{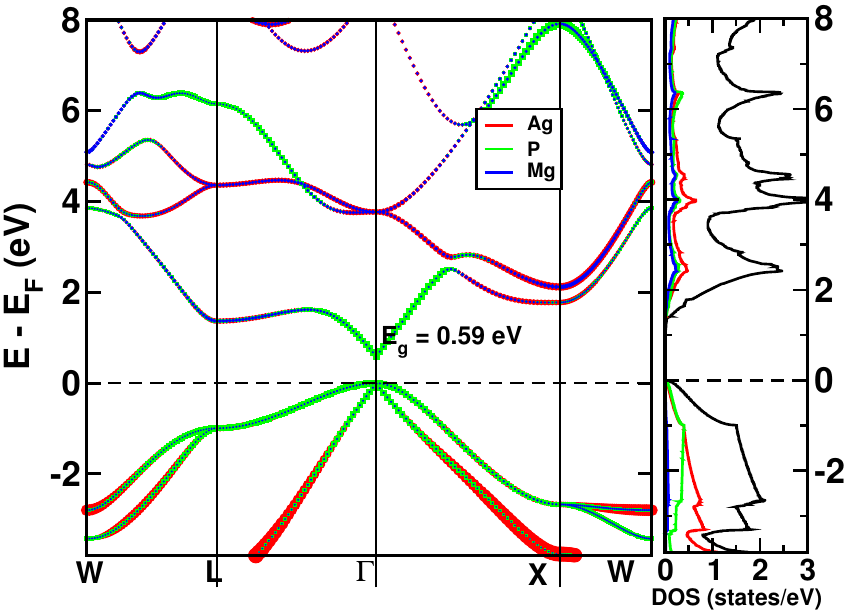}
	\caption{Atom projected electronic band structure and density of states (Fermi level at 0 eV) for AgPMg. }
	\label{bands} 
\end{figure}

\subsubsection{Electronic Structure:}

AgPMg is found to be both chemically ($\triangle E_F$=-368.5 meV/atom) as well as thermally stable with an optimized cubic lattice constant 6.14 {\AA}. The GW band gap is $\sim$ 0.59 eV, which is further used in TE calculations.

Figure \ref{bands} shows the atom projected electronic band structure and the density of states for AgPMg. The valence band edges are mostly composed of Ag while conduction band has contribution from all the three atoms. The valence band edges are relatively more flat than the conduction bands, pointing towards higher hole effective mass. This, in turn, yields a higher Seebeck coefficients for p-type conduction as compared to n-type for a given carrier concentration.

\subsubsection{Phonon Properties:}
\begin{figure}[t]
		\centering
		\includegraphics[scale=0.4]{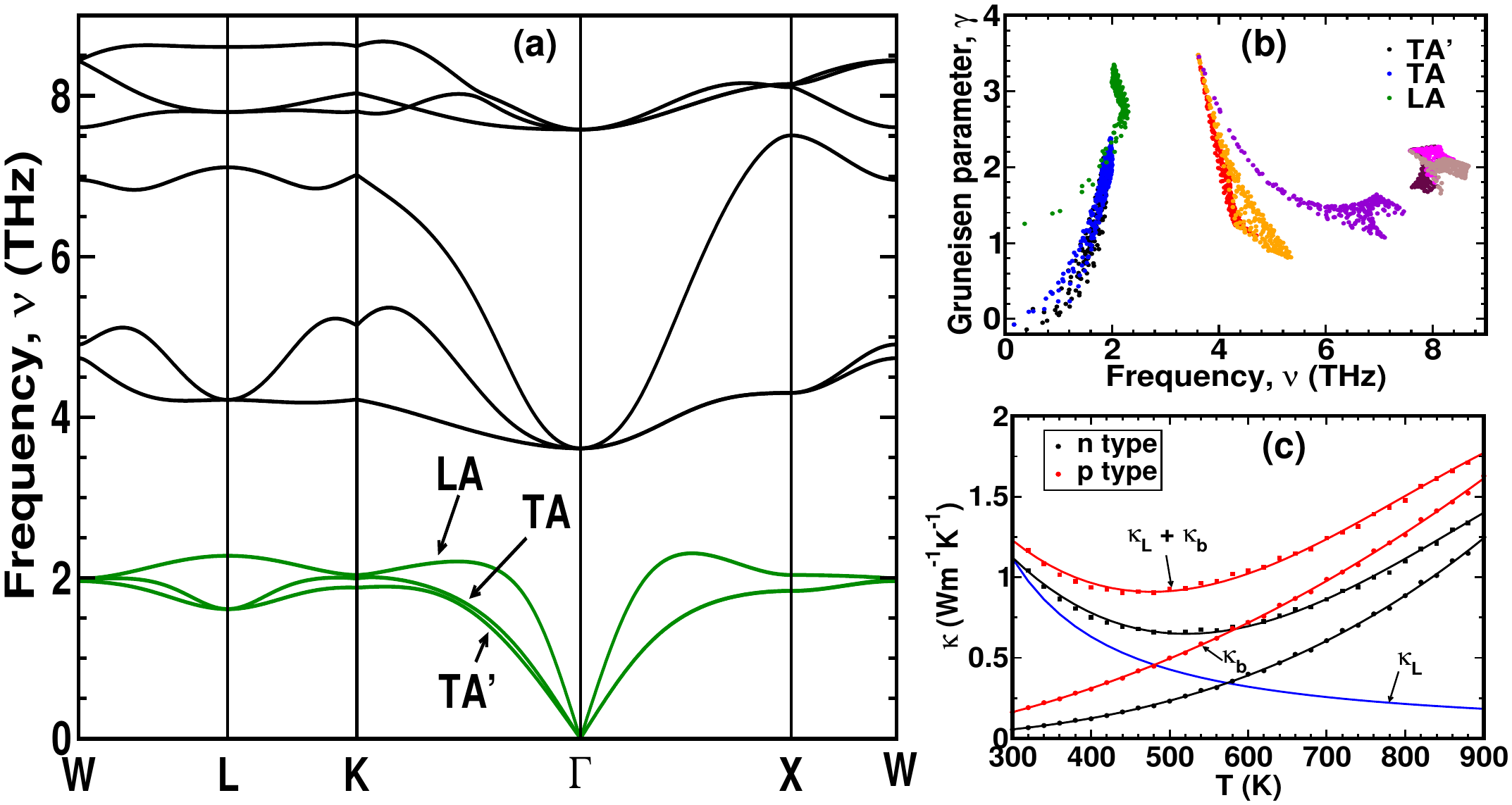}
		\caption{For AgPMg, (a) phonon dispersion,(b) variation of mode Gruneisen parameter $\gamma_i$ with phonon frequency, (c) temperature dependence of lattice and bipolar thermal conductivity (for n and p type carriers).}
		\label{Phonon} 
	\end{figure}

Figure \ref{Phonon}(a) shows the phonon dispersion for AgPMg along high symmetry directions. The acoustic and optical modes of vibration are shown by green and black lines, respectively. Since, the primitive unit cell has three atoms, there are a total of 9 phonon branches (3 acoustics and 6 optical). The acoustic modes are further classified into one longitudinal (LA) and two transverse acoustic modes (TA and TA$^\prime$). The thermal conduction is mainly due to the acoustic modes of vibration. Notably, the two TA and TA$^\prime$ bands are degenerate along $\Gamma$-X and X-W directions. The velocity of each acoustic mode ($\nu$) is given by the slope of the band corresponding to the vibrational mode at the $\Gamma$ point. Debye temperature ($\theta$) can be obtained from the maximum frequency corresponding to the vibrational mode. In addition to $\nu$ and $\theta$, the D-C model for the lattice thermal conductivity also requires the mode Gruneisen parameters ($\gamma_i$). Figure \ref{Phonon}(b) shows the phonon frequency dependence of Gruneisen parameters. The maximum value of $\gamma_i$ for each vibrational mode is taken for the calculation of lattice thermal conductivity in the D-C model. The numerical values of various parameters used to calculate ($k_L$) for AgPMg are; $\nu_{LA}$ = 5170.02 m/s, $\nu_{TA}$ = 2586.92 m/s, $\nu_{TA^{'}}$ = 2296.99 m/s, $\gamma_{LA}$ = 3.35, $\gamma_{TA}$ = 2.38,$\gamma_{TA^{'}}$ = 2.16, $\gamma$=3.35, $\theta_{LA}$ = 110.73 K, $\theta_{TA}$ = 96.44 K, $\theta_{TA^{'}}$ = 94.06 K, V = 19.24 $\times$ $10^{-30}$ $m^{3}$, and M = 54.38 $\times$ $10^{-27}$ kg.

The simulated values of $k_L$ vary between 1.12 to 0.18 W $m^{-1}K^{-1}$ in the temperature range 300-900 K as shown in fig. \ref{Phonon}(c). At sufficiently high temperatures, the effect of minority charge carriers become significant which generates electron-hole pairs and the excitation of electrons across the band gap. 
The thermal transport due to such a process introduces an additional component to $\kappa$, called the bipolar thermal conductivity($\kappa_b$).\cite{bipolar} Figure \ref{Phonon}(c) shows the temperature dependence of $\kappa_b$ for both n-type and p-type carriers. $\kappa_b$ plays an important role in the thermal conductivity beyond  $\sim$ 480 K, as evident from Fig. \ref{Phonon}(c).

\subsubsection{Thermoelectric Properties:}

\begin{figure}[h]
		\centering
		\includegraphics[scale=1.1]{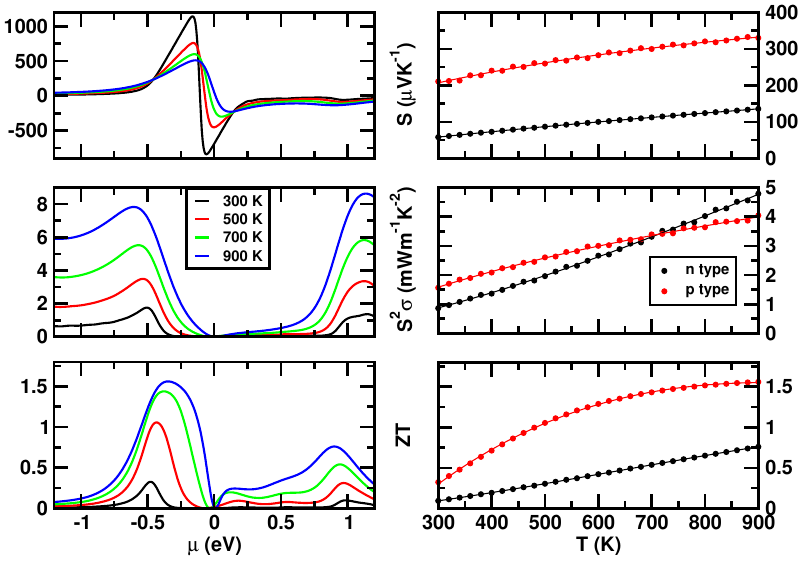}
	\caption{Variation of important TE properties (S, $S^2\sigma$ and ZT) with chemical potential ($\mu$) (left) and temperature (right) for AgPMg. The temperature dependent TE properties were calculated at a carrier concentration of 2.65 $\times$ $10^{20}$ {cm}$^{-3}$ and 5.42 $\times$ $10^{19}$ cm$^{-3}$ for n- and p- type carrier, respectively. }
		\label{TEdata} 
	\end{figure}

\begin{figure}[htbp!]
	\centering
	\includegraphics[scale=1.0]{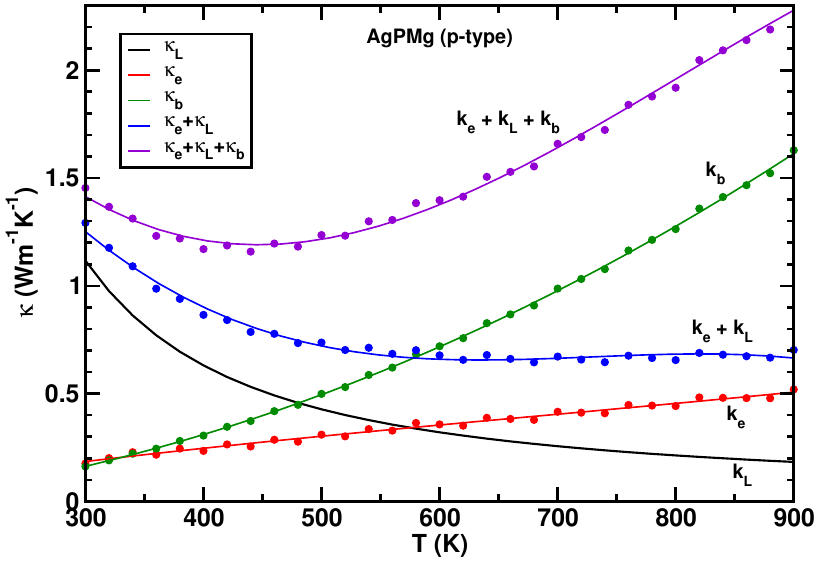}
	\caption{Temperature dependence of lattice ($\kappa_L$), electronic ($\kappa_e$) and bipolar ($\kappa_b$) thermal conductivity for p-type  AgPMg}
	\label{TotalK_ptype} 
\end{figure}

Figure \ref{TEdata} shows the chemical potential ($\mu$) and temperature dependence of the simulated Seebeck coefficient (S), power factor ($S^2\sigma$), and ZT for both n- and p-type conduction for AgPMg.  
The temperature dependent TE properties were calculated for a carrier concentration at which ZT attains its maxmum value. Thes value of these optimized concentrations are 2.65 $\times$ $10^{20}$ ${cm}^{-3}$ and 5.42 $\times$ $10^{19}$ $cm^{-3}$ for n- and p- type conduction, respectively.
At these carrier concentrations, the calculated Seebeck coefficient and power factor for n-type conduction are 135.69 $\mu$V$K^{-1}$ and 4.79 mW$m^{-1}K^{-2}$, whereas for p-type conduction are  337.10 $\mu$V$K^{-1}$ and 3.88 mW$m^{-1}K^{-2}$, respectively. This gives a significantly high value of ZT,  0.76 and 1.56 for n-type and p-type conduction, respectively at 900 K.

To illustrate further, we show, in Figure \ref{TotalK_ptype}, various components of the thermal conductivity with temperature for p-type conduction. The thermoelectric figure of merit is defined as: 
\begin{equation}
\text{ZT} = \frac{S^2\sigma}{(\kappa_e + \kappa_L + \kappa_b)}T
\end{equation}
Since the total thermal conductivity ($\kappa=\kappa_e+\kappa_L+\kappa_b$) is quite small (2.28 W $m^{-1}K^{-1}$) at 900 K (as evident from Fig. \ref{TotalK_ptype}), the ZT value is reasonably  high (1.56) for p-type conduction. ZrNiSn has been shown to be a state of the art material in the HH family for thermoelectric applications. With some selective doping, Ti$_{0.5}$Zr$_{0.25}$Hf$_{0.25}$NiSn, this compound has been proved to be highly promising, with a ZT value of 1.5 at 820 K.\cite{ZrNiSn} In comparison to this state-of-art  material, our proposed compound when lightly doped, exhibits a high thermoelectric figure of merit (ZT$\sim$ 2), making them even better for TE applications. Electronic band structure, phonon dispersion, Gruneisen parameters and temperature dependence of TE properties (S, $S^2\sigma$ and ZT) for the rest of 30 compounds are shown in SM\cite{supl}(Fig SI to Fig SXXX).

\subsection{Solar cell application}

The compounds having band gap in the range 1-1.8 eV were investigated for solar cell application. 
\begin{figure}[t]
	\centering
	\includegraphics[scale=0.65]{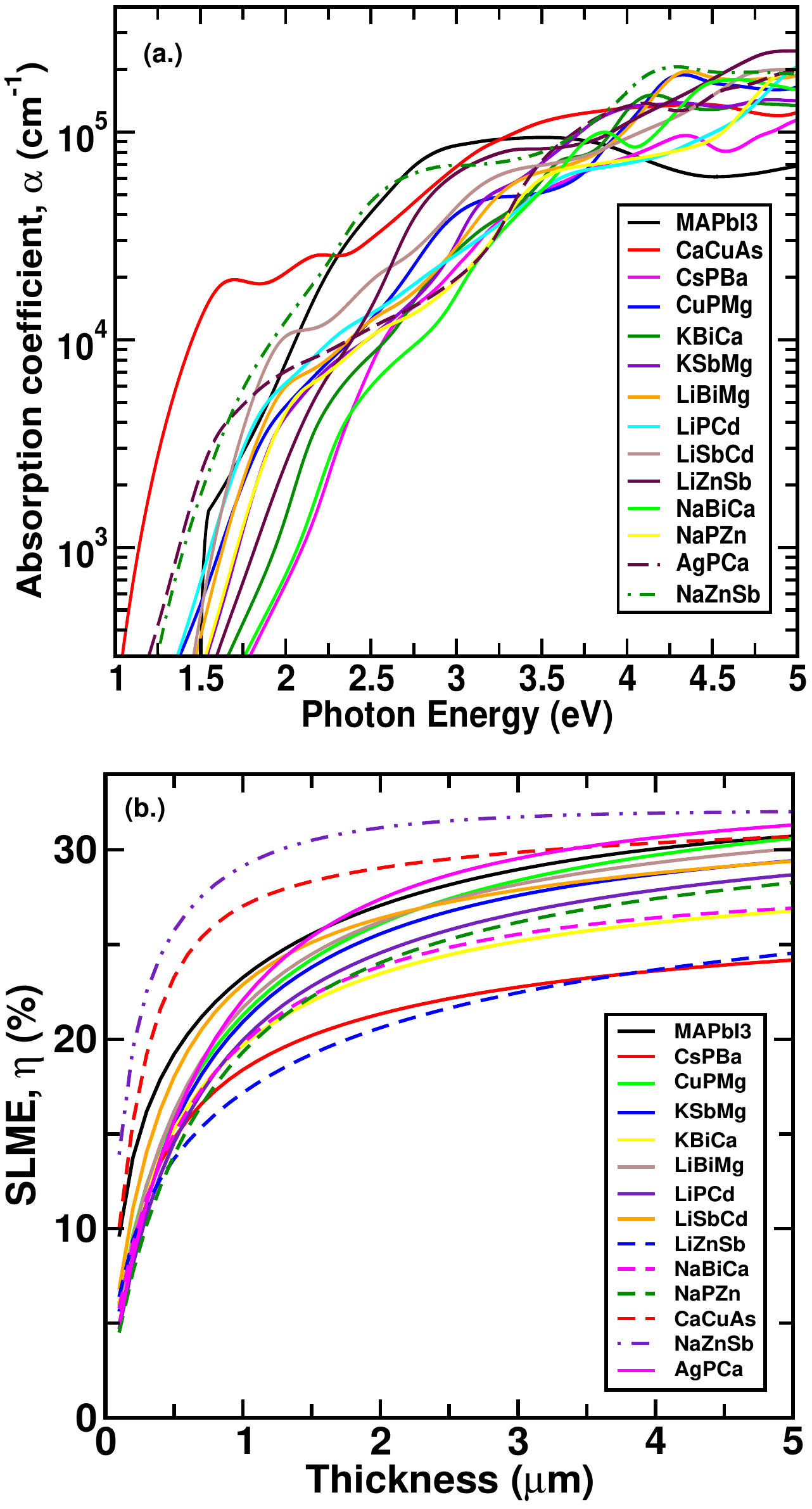}
	\caption{(a) Absorption coefficient vs incident photon energy and (b) "Spectroscopic Limited Maximum Efficiency (SLME)" vs film thickness for compounds having SLME greater than 20$\%$. }
	\label{Abs} 
\end{figure}
First, we simulated the optical transition probabilities for direct transitions from valence band maxima (VBM) to conduction band minima (CBM) for all these compounds. The transition probabilities are estimated by calculating the square of dipole transition matrix elements. Using this, the allowed direct bandgap for all these compounds was calculated. 
The compounds having a difference less than 0.3 eV between optically allowed direct band gap ($E_g^{da}$) and the electronic band gap ($E_g$) were finally shortlisted for further calculations. Out of 30 compounds with electronic band gap between 1-1.8 eV, 18 compounds were found to have a difference, ($E_g^{da}$-$E_g$) $<$ 0.3 eV. We then calculated the next relevant properties, namely the absorption coefficient ($\alpha$) and SLME\cite{solar3}$^{-}$\cite{SLME} for these 18 compounds. The absorption coefficient for a material shows the depth to which a photon with a particular wavelength can penetrate into the material before it gets absorbed. For a given material, calculation of SLME needs specific inputs like the nature/magnitude of bandgap, frequency dependent optical absorption spectra etc.

\begin{figure}[t]
	\centering
	\includegraphics[scale=0.28]{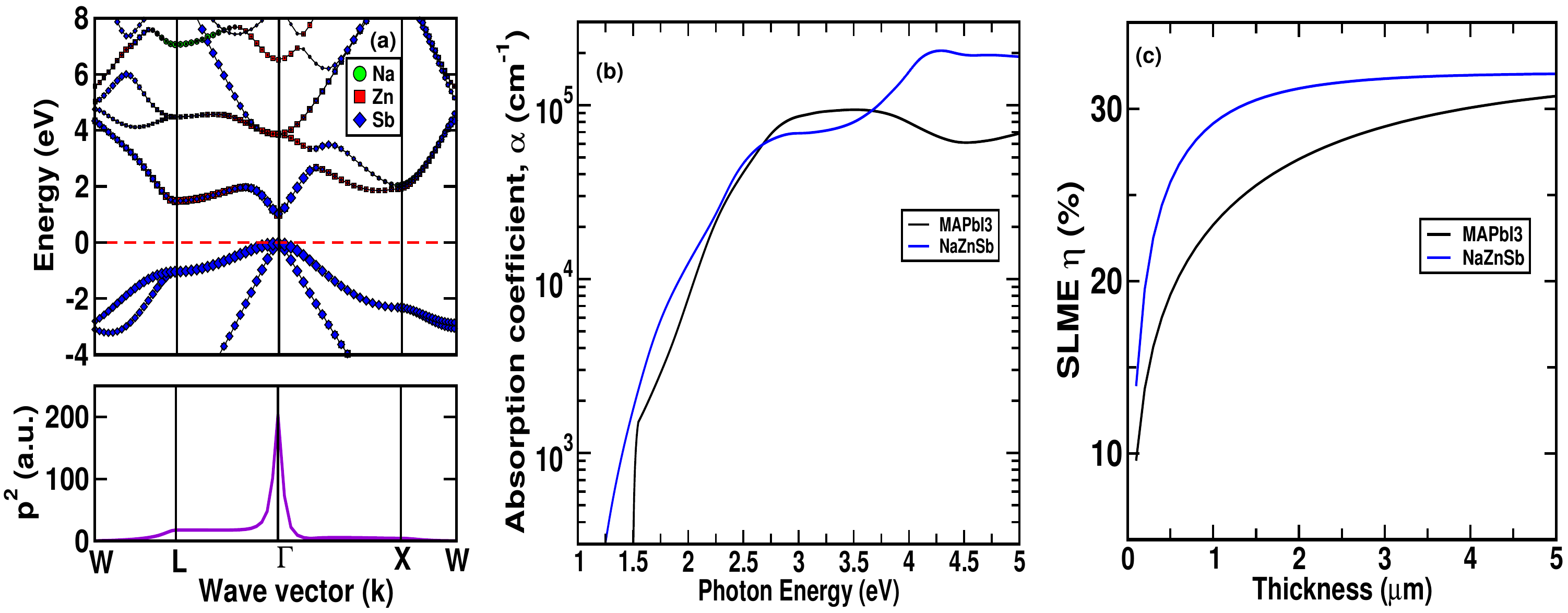}
	\caption{For NaZnSb, (a) Band structure(top) and transition probability (bottom) for NaZnSb. Here green circles show contribution from Na atoms, while red and blue squares represent contribution from Zn and Sb atoms respectively (b) Absorption coefficient vs incident photon energy and (c) "Spectroscopic Limited Maximum Efficiency (SLME)" vs film thickness.}
	\label{NaZnSb} 
\end{figure} 

SLME and GW band gap data for all the 18 compounds is listed in Table SVI of SM.\cite{supl} 13 compounds were found to have SLME greater than 20$\%$. Figure \ref{Abs}(a) shows the frequency dependent absorption coefficient for all the compounds having SLME $>$20$\%$. While Fig. \ref{Abs}(b) shows our calculated SLME vs film thickness compared with the state of the art material $\text{CH}_3\text{NH}_3\text{PbI}_3$ (MAPbI${_3}$).

Interestingly, NaZnSb turns out to be the most promising candidate for solar cell application with SLME $\sim$ 31$\%$ at 2$\mu$m film thickness. Figure \ref{NaZnSb}(a) shows the atom projected band structure and transition probability ($p^2$) for NaZnSb for which HSE - GW band gap and the optimized lattice constant are 1.25 eV and 6.59 $\AA$ respectively. An accurate prediction of band gap is extremely important for the calculation of SLME. As no significant change was observed in the band topology with respect to the exchange correlation functional, we have used PBE band structure with bandgap value shifted to GW values for the calculation of absorption coefficients.  Figure \ref{NaZnSb}(a) (Figure \ref{NaZnSb}(b)) shows a comparison of the optical absorption coefficient (SLME) for NaZnSb and MAPbI${_3}$. One can notice that, absorption coefficient of NaZnSb is overall larger compared to those of MAPbI$_3$. SLME for the proposed  compound is also much higher, approaching the Shockley–Queisser limit, as compared to MAPbI${_3}$. Also, this value is comparable to the efficiency potential of the GaAs/Si dual-junction solar cell which has been reported to be 31.7\% in previous study.\cite{Tandem_solar} Electronic band structure, transition probability and phonon dispersion for the rest of 12 compounds are shown in SM\cite{supl}(see Fig. SXXX to SXLIV).

\subsection{Topological Insulators}
\begin{figure}[h!]
	\centering
	\includegraphics[scale=1.0]{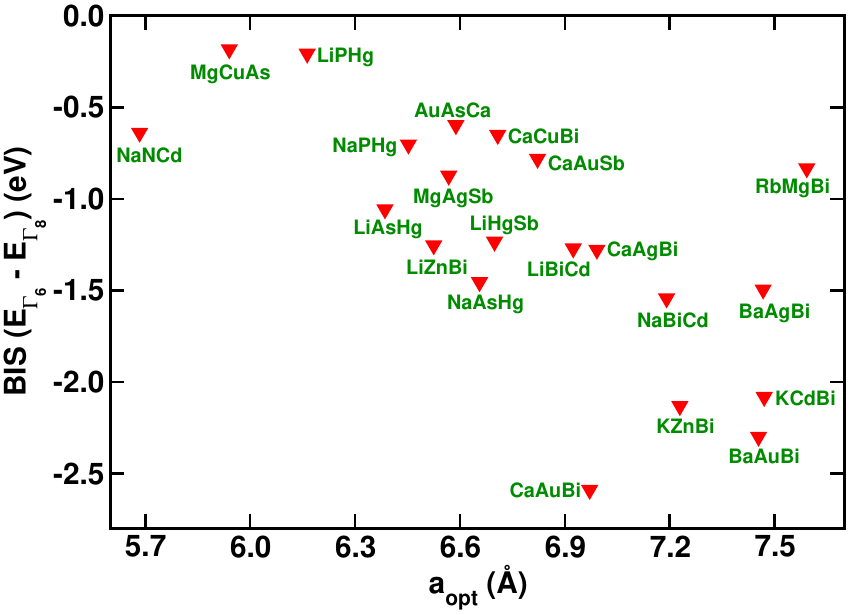}
	\caption{The band inversion strength, BIS ({$\text{E}_{\Gamma_6}$}-{$\text{E}_{\Gamma_8}$}) vs optimized lattice constants ($a_{opt}$) for all the topologically non-trivial compounds in the 8-electron HH family.}
	\label{BIS} 
\end{figure}

Topological quantum materials are the other class of potential materials which often find home in half Heusler alloys. These materials are usually zero (small) band gap systems with surface protected bulk band inversion. While screening the 8-electron half Heusler compounds, we found several of them having semi-metallic and/or weakly metallic behavior where the conduction bands get mixed with fewer valence bands (within the PBE functional). Topological insulators have non-trivial band topology induced by spin-orbit coupling which requires inverted band order between s-like ($\Gamma_6$ band) and p-like ($\Gamma_8$ band) orbitals at the $\Gamma$ point. Band Inversion strength (BIS) is a measure of the non triviality which is defined as the difference between the energy of s-like ($\Gamma_6$ band) and p-like ($\Gamma_8$ band) bands at the $\Gamma$ point. The negative value of BIS implies the non-trivial band order. Because the band topology and the magnitude of BIS is extremely sensitive to the computational method used in a given study, we have performed HSE-SOC band structure calculations for all those compounds which are semimetallic or weakly metallic at the PBE level. Figure \ref{BIS} shows the BIS and optimized lattice parameters of 21 compounds which display non-trivial band topology at ambient conditions.  
It should also be noted that, for compounds having small positive BIS values (e.g. MgCuSb having BIS value 0.37), the band inversion can be induced by applying small perturbation such as pressure, strain, or doping.
\begin{figure}[h!]
	\centering
	\includegraphics[scale=1.2]{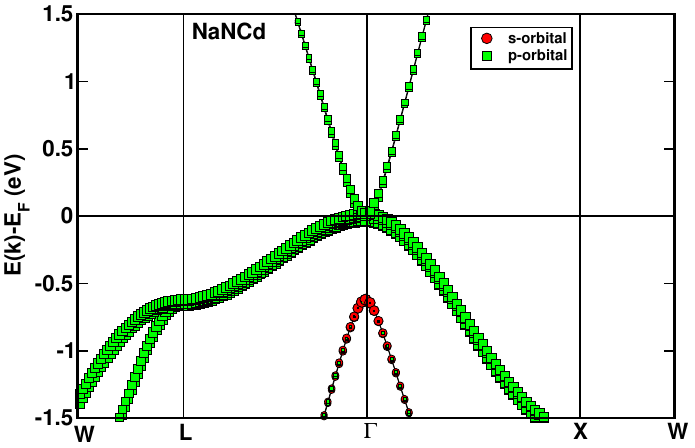}
	\caption{HSE06+SOC bulk bandstructure of NaNCd at ambient lattice parameter (a = 5.68 \AA). Red circles and green squares represent the s and p orbital character respectively.}
	\label{Bulk} 
\end{figure}
  
\begin{figure}[h!]
	\centering
	\includegraphics[scale=0.7]{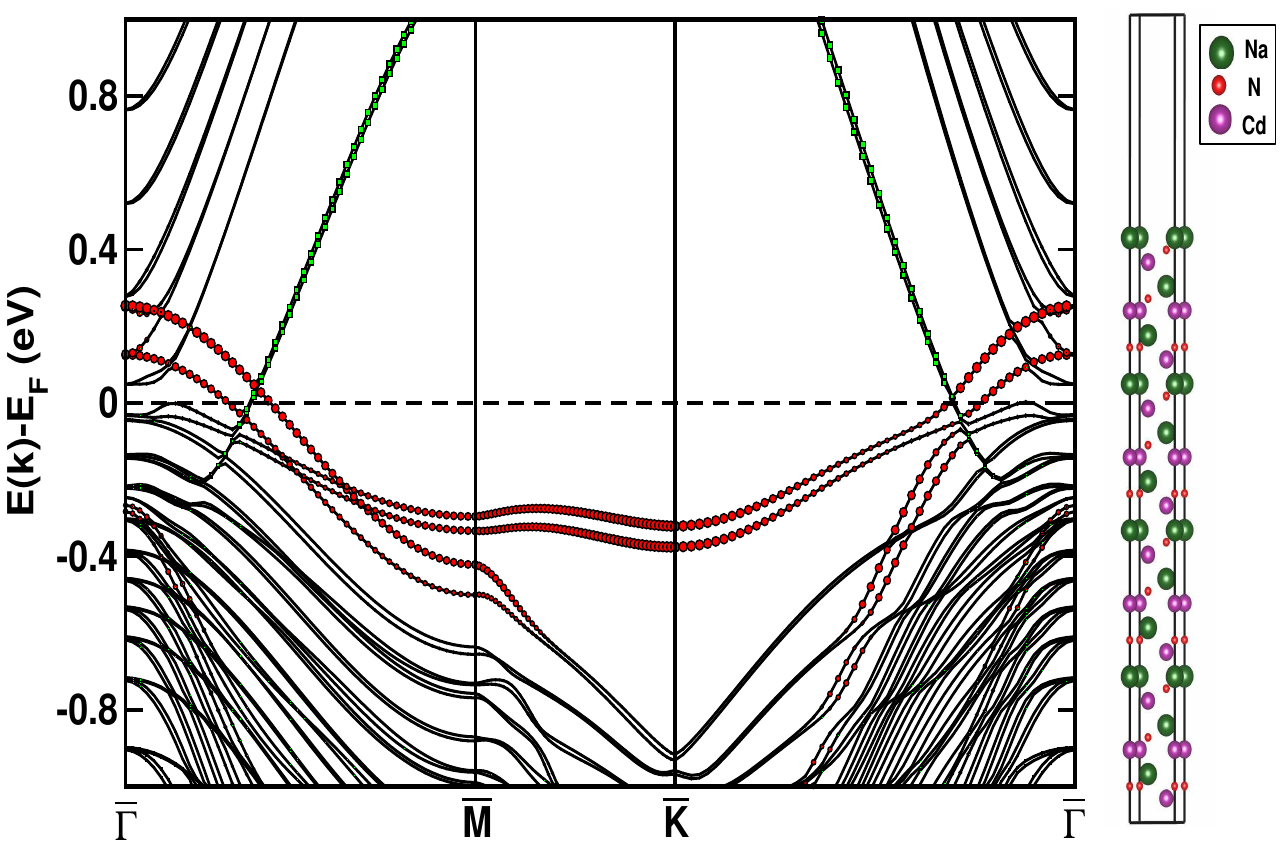}
	\caption{(Left)Surface bands for NaNCd (red circles and green squares represent the contribution of surface atoms from the bottom three (X-Y-Z) and top three (Y-Z-X) atomic layers, respectively). (Right) unit cell for (111) surface with Na termination at the top and Cd at the bottom.}
	\label{surface} 
\end{figure}

For TIs, surface states play a crucial role in further confirming the topological behavior of the material. Here, we chose a representative system NaNCd with BIS value -0.64 to further study the surface properties. Figure \ref{Bulk} shows the bulk band structure of NaNCd at optimized lattice constant (a = 5.68 $\AA$) where red circles and green squares represent s and p-orbital contributions, respectively. As seen from the figure, p-like bands have four-fold degeneracy and s-like states lie below them. The band inversion is just the necessary condition whereas the robustness of conducting states is a sufficient condition for an acceptable TI.\cite{topo}$^{,}$\cite{topo1} 

HH compounds have a layered structure along the [111] direction. Thus, the most naturally cleavable surface is along the [111] plane. Hence, for investigating surface bands, a surface slab along [111] direction was constructed for NaNCd at the optimized lattice parameter. A 36 atomic-layer thick slab (38.02$\AA$), with Na atom termination at top and Cd termination at the bottom surface, is shown in Fig. \ref{surface}. A vacuum of 15 $\AA$ was used to make sure there is no interaction between the repeating surface layers due to the periodic boundary conditions. Five atomic layers on each of the top and bottom sides of the slab was fully relaxed, keeping the remaining middle layers fixed, such that force (energy) is converged up to 0.001 eV/$\AA$ ($10^{-6}$ eV).

\begin{table}[htbp!]
	\centering{}
	\begin{tabular}{|c|c|c|}
		\hline 
	 Slabs  & Surface energy(eV/$\AA$\textsuperscript{2})    \tabularnewline
	\hline
	\hline
	 Na (top surface)    & 0.028  \\
	 and Cd(bottom surface) terminated    & \\
	\hline
	 N (top surface)    &0.037  \\
	 and Cd(bottom surface) terminated  & \\
	\hline
	 Both side Cd terminated    & 0.1352     \\
	
	\hline

	\hline
	        \end{tabular}                                                                                    
	
	\caption{Surface energies of various terminated surfaces for surface slab of NaNCd along [111] direction. }   \label{Surface energies}
\end{table}

 Table \ref{Surface energies} lists the surface energies of the slab with different termination layers. The slab with Na and Cd termination at the top and bottom surfaces respectively is found to be energetically the most favorable surface and hence our choice for further simulations. The surface band structure for this slab is shown in Fig. \ref{surface}. The size of red circles and green squares represent the weighted contributions of surface atoms from bottom three and top three consecutive atomic layers, respectively. Clearly the surface bands show metallic character and hence conducting in nature. HSE-SOC structures showing the band inversion, for the rest of 20 compounds are shown in SM\cite{supl} (see Fig. SXLV to SLIV).

\section{Conclusion}

To summarize, our work is focused on high throughput calculation of I-II-V class of compounds for searching promising materials for renewable energy applications. 121 out of 960 compounds were found to be both chemically and thermally stable. An accurate HSE-GW level band gap calculation lists out 31 compounds for thermoelectric application having band gap in the range 0-1.5 eV. 30 compounds having band gap in the range 1-1.8 eV were selected to study solar harvesting problem. 22 compounds which were metals/semi-metals were checked for band inversion in presence of spin-orbit coupling for identifying potential topological insulators. 13 compounds were found to have thermoelectric figure of merit (ZT value) greater than equal to 0.7. 13 compounds show excellent optoelectronic properties with SLME greater than 20$\%$. 21 compounds were found to show band inversion for which robustness of surface conducting states was checked to make them qualify as strong topological insulators. The properties of proposed compounds from our calculations are found to be comparable to the state-of-art materials. We believe our detailed high throughput investigation and the accurate shortlisting of potential materials will be of timely interest to the readers and should be helpful in recent pursuit of novel functional materials for renewable energy applications.  We strongly suggest experimentalists to synthesize these shortlisted compounds for corresponding applications and verify our claim. 






\section*{Acknowledgment}
Bhawna acknowledges financial support from Indian Institute of Technology, Bombay in the form of teaching assistantship. 
AA acknowledges National Centre for Photovoltaic Research and Education (NCPRE) (financially supported by Ministry of new renewable energy 
(MNRE), Government of India) to support this research.

\end{document}